\documentclass[preprint,onecolumn,nofootinbib]{revtex4}
\pdfoutput=1
\usepackage[colorlinks=true,linkcolor=blue,urlcolor=blue,filecolor=black,citecolor=red,pdfstartview=FitV,pdftitle={},pdfsubject={},pdfkeywords={},pdfpagemode=None,bookmarksopen=true]{hyperref}
\usepackage{graphicx}
\usepackage{amsmath}
\usepackage{amsfonts}
\usepackage{amssymb,ulem}
\usepackage{color}%
\usepackage{CJK}
\usepackage{dcolumn}
\newcommand{\f}{\begin{equation}}
\newcommand{\ff}{\end{equation}}
\newcommand{\fa}{\begin{eqnarray}}
\newcommand{\ffa}{\end{eqnarray}}

\begin{document}
\title{AC charge transport in holographic Horndeski gravity}
\author{Xi-Jing Wang$^{1}$}
\thanks{xijingwang@mail.dlut.edu.cn}
\author{Hai-Shan Liu$^{2,3}$}
\thanks{hsliu.zju@gmail.com}
\author{Wei-Jia Li$^{1}$}
\thanks{weijiali@dlut.edu.cn}
\affiliation{
$^1$ Institute of Theoretical Physics, School of Physics, Dalian
University of Technology, Dalian 116024, China\ \\
$^2$ Center for Joint Quantum Studies,Tianjin University, Tianjin 300350, China\ \\
$^3$ Institute for Advanced Physics and Mathematics, Zhejiang University of Technology, Hangzhou 310023, China}
\begin{abstract}
In this paper, we investigate the AC charge transport in the holographic Horndeski gravity and identify a metal-semiconductor like
transition that is driven by the Horndeski coupling. Moreover, we fit our numeric data by the Drude formula in slow relaxation cases.
\end{abstract}
\maketitle
\tableofcontents
\section{Introduction}
Understanding the dynamics of various exotic quantum matters, including quark-gluon plasma as well as strongly coupled electronic systems, has long been a core topic in modern physics \cite{Hartnoll:0903,McGreevy:1606,Hartnoll:1612,Baggioli:1908}. These systems are, in general, hard to deal with by conventional perturbative theories due to the existence of strong interactions between the microscopic degrees of freedom. The string theory inspired holographic duality, as a strong/weak duality, offers a powerful tool to touch the anomalous phenomena in strongly coupled systems, such as the linear dependance of DC resistivity on temperature in a wide class of strange metals.

Physically, symmetry of spatial translation implies the conservation law of momentum. For finite density system without momentum dissipations, the DC conductivity is divergent because of a non-trivial overlap between the electric current and the conserved momentum. However, in most of realistic materials, such a symmetry is  broken for the presence of lattice, impurities or defect, which results in the finite conductivity. Then, realizing effects of momentum dissipation in holography has been extensively studied in recent years \cite{Horowitz:1204,Horowitz:1209,Horowitz:1302,Donos:1311,Ling:1410,Ge:1606,Ge:1412,Cremonini:1608,Chesler:1308,Gouteraux:1401,Rangamani:1505,Ling:1309,Donos:1409,Andrade:1311,WJL:1602,Davison:1306,Blake:1308,Baggioli:1411,Alberte:17082}.

In this paper, we shall study the holographic charge transport of Horndeski gravity model. The Horndeski gravity was first constructed in 1970s \cite{Horndeski:1974,Kobayashi:1901}, and was recently rediscovered in cosmology \cite{Nicolis:0811,gaoxian}. The theory has a remarkable property that it contains higher than two derivatives at the level of Lagrangian, however, each field has at most two derivatives at the level of equation of motion. This property is similar to that of Lovelock gravity \cite{Lovelock:1971}, and thus the theory can be ghost free. The black hole solutions of the theory  were constructed in \cite{Anabalon:1312,Cisterna:1401} and their thermodynamics were carried out in \cite{Feng:1509,Feng:1512}. The stability and causality were studied in \cite{Beltran:1308,Kobayashi:1402,Minamitsuji:2015}. It was showed in \cite{atheorem} that the theory admits the holographic a-theorem in a critical point of the parameter space, which suggests the theory may have a holographic dual field theory. Further holographic properties and applications of Horndeski theory were investigated in \cite{Kuang:1607,Caceres:1710,Feng:1811,Li:1812}. The DC limit of holographic transports of Horndeski gravity models were extensively analytically studied in \cite{HSL:1703,WJL:1705}, especially it was pointed out that the thermal conductivity bound \cite{Grozdanov:1511} was broken by a type of Horndeski model \cite{HSL:1804}. Here, we want to go a step further to study the properties of the AC electric conductivity of the Horndeski theory through numerical methods.


We first briefly go over a simple holographic Horndeski model with momentum relaxation which is the Einstein-Maxwell sector coupled with an axion-Einstein tensor, its black hole solutions and the result of DC conductivity \cite{HSL:1703}. Then, by introducing time-dependent perturbations, we derive linearized equations of motion. And, the AC electric conductivity can be read from the near boundary behavior after solving these bulk equations in a  numeric approach. The dependance of the AC conductivity on frequency are displayed for different temperatures and Horndeski couplings. The numeric results imply that this holographic model could provide a mechanism of semiconductor-metal transition. Moreover, in the slow relaxation region, the conductivity seems perfectly fitted by Drude formula as follows,
\begin{equation}
\sigma(\omega)=\frac{\sigma_{DC}}{1-i\omega\tau_{rel.}},
\end{equation}
where $\tau_{rel.}$ is the relaxation time, which indicates the speed of momentum relaxation and can also be calculated by using the hydrodynamical method \cite{Davison:1306}.

This paper is organized as follows: In section II, we introduce the black hole background in the Horndeski theory. In section III, we calculate the electric conductivity by turning linearized perturbations of the bulk fields and show the numeric results. In section IV, we conclude. Finally, we show the details of calculating AC conductivity in the appendix.

\section{Holographic setup}
In this paper, we consider the following holographic action \cite{HSL:1703}
\begin{equation}
S= \int\,d^4 x \sqrt{-g}\left(\kappa(R-2\Lambda-\frac{1}{4}F^2)-\frac{1}{2}(\alpha g^{\mu\nu}-\gamma G^{\mu\nu}) \sum_{i=1}^2\partial_\mu \phi^i\partial_\nu \phi^i\right), \label{action}
\end{equation}
where $\Lambda$ is the cosmological constant, $G^{\mu\nu}\equiv R^{\mu\nu}-\frac{1}{2}Rg^{\mu\nu}$ is the Einstein tensor, $F_{\mu\nu}\equiv\nabla_\mu A_\nu-\nabla_\nu A_\mu$ is the electromagnetic field tensor and $\kappa$, $\alpha$, $\gamma$ are coupling constants. As is mentioned in the introduction, there exists a critical value, $\alpha + \gamma \Lambda = 0$, in the parameter space, where theory admits holographic $a$-theorem \cite{atheorem}.   For simplicity, we shall set $\kappa=\alpha=1$ from now on. $\phi^i$ are massless axion fields which are coupled to the Einstein tensor. Thus $\gamma$ should be regarded as the Horndeski coupling strength. If we set $\gamma=0$, it returns to Einstein-Maxwell theory with two free axions\,\cite{Andrade:1311}.
From the above action (\ref{action}), the Einstein equation, Klein-Golden equation as well as Maxwell equation are given by
\begin{align}
& G_{\mu\nu}+\Lambda g_{\mu\nu}-\frac{1}{2}F_{\mu\rho}{F^{\rho}}_\nu+\frac{1}{8}F^2 g_{\mu\nu}-\frac{1}{2}\sum_{i=1}^2\left(\partial_\mu \phi^i\partial_\nu \phi^i-\frac{1}{2}g_{\mu\nu}\left(\partial \phi^i\right)^2\right)\nonumber\\
&-\sum_{i=1}^2\frac{\gamma}{2}\Big\{\frac{1}{2}\partial_\mu \phi^i\partial_\nu \phi^i R-2\partial_\rho \phi^i{\partial_{(\mu}\phi^i R_{\nu)}}^\rho-\partial_\rho\phi^i\partial_\sigma\phi^i{{{R_\mu}^\rho}_\nu}^\sigma\nonumber\\
&-\left(\nabla_\mu\nabla^\rho\phi^i\right)\left(\nabla_\nu\nabla_\rho\phi^i\right)+\left(\nabla_\mu\nabla_\nu\phi^i\right)\Box\phi^i+\frac{1}{2} G_{\mu\nu}\left(\partial\phi^i\right)^2\nonumber\\
&-g_{\mu\nu}\left[-\frac{1}{2}\left(\nabla^\rho\nabla^\sigma\phi^i\right)\left(\nabla_\rho\nabla_\sigma\phi^i\right)+\frac{1}{2}\left(\Box \phi^i\right)^2-\partial_\rho\phi^i\partial_\sigma\phi^i R^{\rho\sigma}\right]\Big\}=0,\\[0.2cm]
&\nabla_\mu\left[\left(g^{\mu\nu}-\gamma G^{\mu\nu}\right)\partial_\nu\phi^i\right]=0,\qquad \nabla_\nu F^{\nu\mu}=0.
\end{align}
To avoid the ghost problem, the Horndeski coupling constant $\gamma$ should satisfy \cite{HSL:1703}
\begin{gather}
-\infty<\gamma\le-\frac{1}{\Lambda}.
\end{gather}
We consider the static ansatz
\begin{gather}
ds^2=\frac{1}{u^2}\bigg[-h(u)dt^2+\frac{1}{f(u)}du^2+dx^idx^i\bigg],\\
A=A_t(u)dt,\qquad \phi^i=k \,x^i,
\end{gather}
where $k$ is a constant, and $x^i$ denotes the boundary spatial coordinates. In our convention, the AdS boundary is located at $u=0$. The Horndeski axion fields provide the momentum relaxation in the dual boundary field theories. The Einstein and Maxwell equations then turn out to be
\begin{gather}
f u^4 A_t'{}^2-4 h u f'+2 h \left(-\gamma  f k^2 u^2+6 f+k^2 u^2+2 \Lambda \right)=0,\\
f u^4 A_t'{}^2-4 f u h'+2 h \left(\gamma  f k^2 u^2+6 f+k^2 u^2+2 \Lambda \right)=0,\\
h u f' \left(u h'-h \left(\gamma  k^2 u^2+4\right)\right)+2 h^2 \left(f \left(\gamma  k^2 u^2+6\right)+2 \Lambda \right)\notag\\-f u \left(h u \left(u^2 A_t'{}^2-2 h''\right)+h h' \left(\gamma  k^2 u^2+4\right)+u h'^2\right)=0,\\
A_t' \left(f h'-h f'\right)-2 f h A_t''=0.
\end{gather}
These equations can be solved analytically
\begin{gather}
h(u)=U(u)f(u),\qquad U(u)=e^{\frac{ \gamma  k^2 u^2}{2}},\\
A_t(u)=\mu -\frac{\sqrt{\pi }\, q\, \text{erfi}\left(\frac{ \sqrt{\gamma } k u}{2}\right)}{\sqrt{\gamma }\, k},\\
f(u)=\frac{e^{-\frac{\gamma  k^2 u^2}{4} }}{12 \sqrt{\gamma } k u_h^3}\bigg(\sqrt{\pi } u^3 u_h^3 \left(\gamma  k^4 (\gamma  \Lambda +3)+3 q^2\right) \big(\text{erfi}(\frac{ \sqrt{\gamma } k u}{2})-\text{erfi}(\frac{ \sqrt{\gamma } k u_h}{2})\big)\notag\\+2 \sqrt{\gamma } k \big(u^3 e^{\frac{\gamma  k^2 u_h^2}{4} } \left(k^2 (\gamma  \Lambda +3) u_h^2+2 \Lambda \right)-u_h^3 e^{\frac{\gamma  k^2 u^2}{4} } \left(k^2 u^2 (\gamma  \Lambda +3)+2 \Lambda \right)\big)\bigg),
\end{gather}
where erfi(x) is the imaginary error function, defined by erfi(x)=$\frac{2}{\sqrt\pi}\int_0^x e^{t^2}\,dt$. It is easy to prove that when setting $\gamma=0$ the black hole solutions in  \cite{Andrade:1311} are recovered. Here, $u_h$ is the event horizon which is determined by $f(u_h)=0$. $\mu$ and $q$ are integration constants which should be identified as the chemical potential and the charge density of the dual field theory, respectively. On the horizon, the regularity of the gauge field $A_\mu$ requires that these two quantities should obey the following relation
\begin{gather}
\mu=q\frac{\sqrt{\pi }  \,\text{erfi}\left(\frac{ \sqrt{\gamma } \,k\, u_h}{2}\right)}{\sqrt{\gamma }\, k}\label{1}.
\end{gather}
Finally, the Hawking temperature is given by
\begin{gather}
T=\frac{\sqrt{f'(u_h)h'(u_h)}}{4\pi}=-\frac{e^{\frac{ \gamma\,  k^2 \,u_h^2}{4}} \left(2 k^2\, u_h^2+q^2 \,u_h^4+4 \,\Lambda \right)}{16\, \pi\,  u_h}\label{2}.
\end{gather}
\section{Electric conductivity}
The DC conductivity can be calculated analytically through various methods based on the "membrane paradigm" \cite{HSL:1703,Blake:1308,Donos:1406,Amoretti:1407,Iqbal:0809,Donos:1506}. The pivotal point of these methods is that one constructs a radially conserved current which connects the horizon and the boundary. Thus, the DC conductivity of the boundary system can be expressed in terms of the horizon data as follows \cite{WJL:1705}
\begin{gather}
\vspace{1cm}
\sigma_{DC}=1+\frac{q^2}{M_h^2}\label{4},
\vspace{1cm}
\end{gather}
where $M_h^2$ is the effective graviton mass on the horizon given by
\begin{gather}
\vspace{1cm}
M_h^2=\frac{k^2}{u_h^2}-\gamma\left(\frac{4\pi\, k^2\, T\,e^{-\frac{ \gamma k^2 u_h^2}{4}}}{u_h}\right)\label{5}.
\vspace{1cm}
\end{gather}
\begin{figure}[htbp]
  \vspace{0cm}
  \centering
  \includegraphics[width=0.55\textwidth]{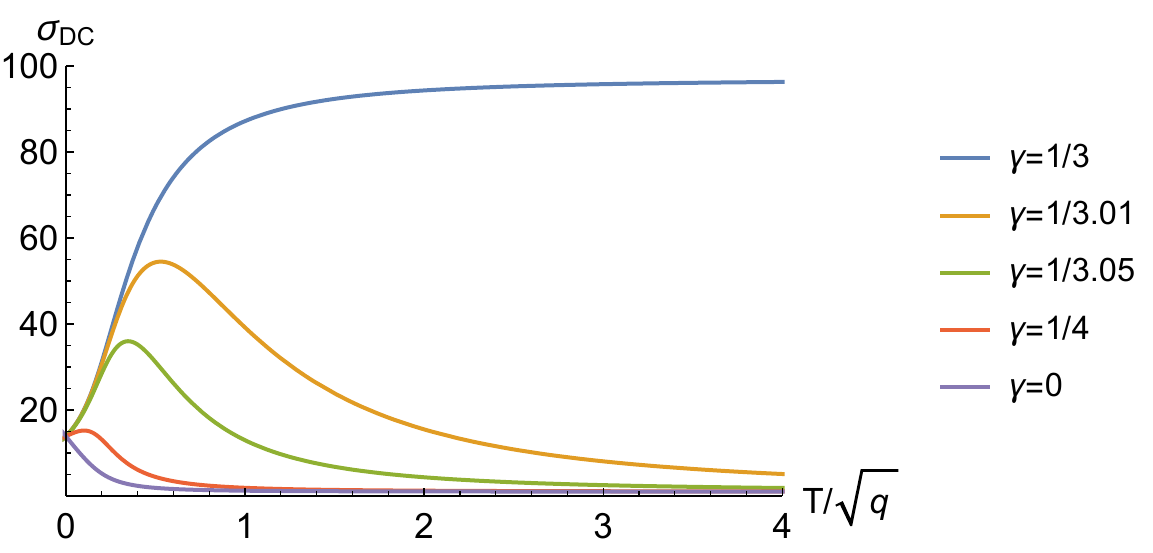}
  \caption{DC conductivity $\sigma_{DC}$ as the function of temperature $T$ for different Horndeski couplings $\gamma$, where we set $k=1/2$ and $\Lambda=-3$. And, there is a critical value which is $\gamma=1/3$ beyond which the conductivity monotonously increases as the increases of the temperature. This figure has been firstly worked out in \cite{HSL:1703}.}\label{a}
  \vspace{0cm}
\end{figure}

\begin{figure}[htbp]
  \vspace{0cm}
  \centering
  \includegraphics[width=0.46\textwidth]{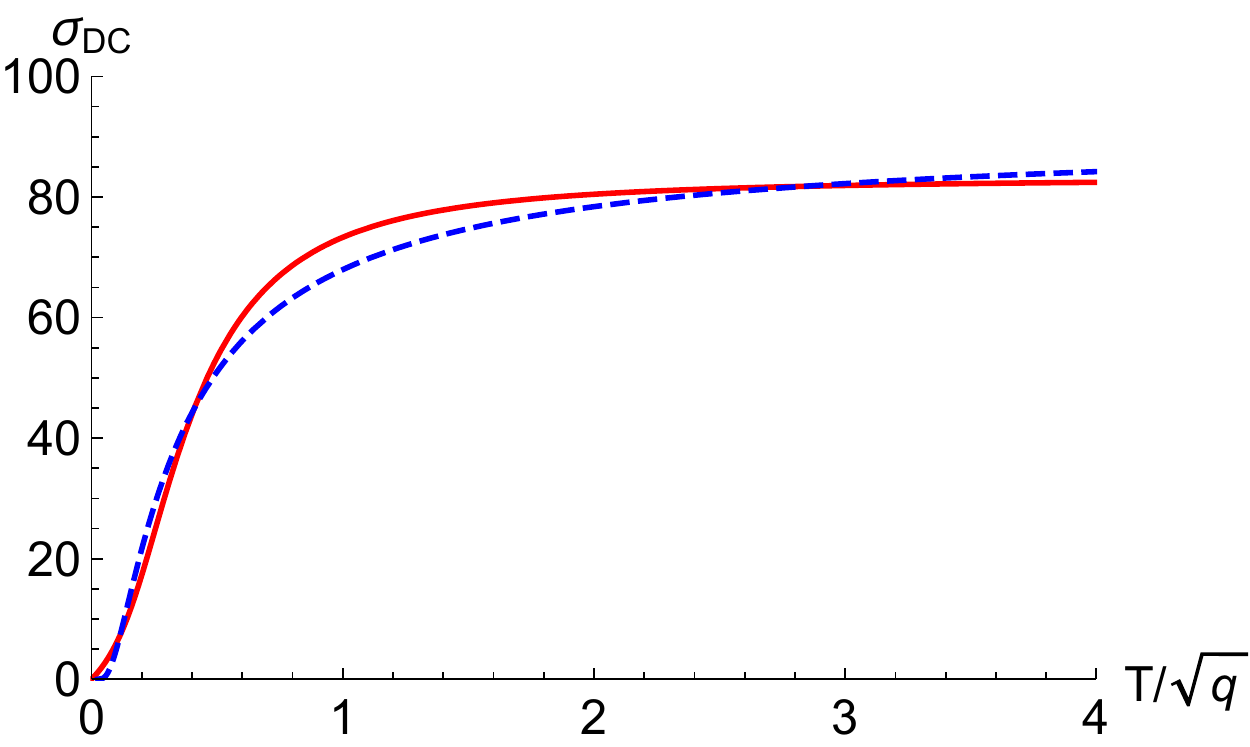}
  \caption{Red solid line is  the result of holographic theory without the zero temperature conductivity contribution obtained from (\ref{4}) and $\gamma$ is at the critical value $1/3$. Blue dashed line is the result obtained from (\ref{z}). We have fixed $k=1/2$ here. Coefficients $A=15.7127,\,B=3.48369$,\, and $C=0.0002147$.}\label{a1}
  \vspace{0cm}
\end{figure}

For convenience, we set $q=1$, $k=1/2$, and $\Lambda=-3$ from now on. Thus, $\gamma$ is equal $1/3$ at the critical value. The figure of DC conductivity $\sigma_{DC}$ with temperature $T$ is shown in FIG.\ref{a}. As seen in FIG.\ref{a}, when the Horndeski coupling is vanishing, $\gamma=0$, the conductivity declines monotonically from a non-zero initial value as the temperature increases, behaving like a metal. When $\gamma$ is at the critical value $1/3$, the conductivity increases monotonically from non-zero initial value as the temperature increases behaving like a semiconductor. When $\gamma$ is below the critical value, the conductivity firstly increases and then declines as the temperature goes up. Note that for most of semiconductors, their conductivities obeys the Steinhart-Hart equation\,\cite{SH}
\begin{gather}
\frac{1}{T}=A+B\,\text{log}\,\rho+C\,(\text{log}\,\rho)^3,\label{z}
\end{gather}
where $\rho$ is resistivity (the inverse of conductivity), and $A$,\,$B$,\,and $C$ are so-called Steinhart-Hart coefficients to be determined. We can find that the DC conductivity of our model fits this equation quite well by subtracting the universal unit part which comes from the absence of  Galilean invariance in critical systems. Thus in order to analyze the problem qualitatively and fit the conductivity curve, the zero temperature conductivity contribution should be ignored. We draw the figure of conductivity with temperature by using formula (\ref{z}) and show the comparison with the conductivity curve of the critical coupling $\gamma=1/3$ in FIG.\ref{a1}. According to this figure, the behaviors of these two conductivities are in qualitative agreement. In other words, when $\gamma$ is at the critical value, the conductivity increases monotonically with the temperature increasing, behaving like a semiconductor. Then, it seems that there exists a semiconductor-metal phase transition driven by the Horndeski coupling term at some certain critical temperature $T_c$ when $\gamma$ is below the critical value. On top of this, we will further explore how the AC conductivity behaves in both phases.

We follow the procedure illustrated in \cite{Andrade:1311} to calculate AC conductivity. Take the following forms of the time-dependent linearized perturbations
\begin{gather}
\delta g_{tx}=e^{-i\omega t}\frac{h_{tx}(u)}{u^2},\qquad  \delta A_x=e^{-i\omega t}a_x(u),\qquad\delta\phi^x=e^{-i\omega t}\frac{\chi(u)}{k}.
\end{gather}
The linearized equations of motion of metric field\,(\,the $(u,\,x)$ component), gauge field and scalar field are given by
\begin{gather}
i q u^2 \omega  a_x e^{\frac{1}{4} \gamma  k^2 u^2}+\chi ' \left(\gamma  f u h'-3 \gamma  f h+h\right)-i \gamma  k^2 u \omega  h_{tx}-i \omega  h_{tx}'+\gamma  u \chi  \omega ^2=0\label{6},\\
a_x' (h f'+\frac{1}{2} \gamma  f h k^2 u)+f h a_x''+\omega ^2 a_x-f q h_{tx}' e^{\frac{1}{4} \gamma  k^2 u^2}=0\label{7},\\
\chi '' \left(\gamma  f^2 u h'-3 \gamma  f^2 h+f h\right)+h_{tx} \left(-i \gamma  k^2 u \omega  f'+3 i \gamma  f k^2 \omega -i k^2 \omega \right)+\chi  \left(\gamma  u \omega ^2 f'-3 \gamma  f \omega ^2+\omega ^2\right)\notag\\+\chi ' (\gamma  f^2 u h''-\frac{5}{2} \gamma  f^2 h'-\frac{\gamma  f^2 u \left(h'\right)^2}{2 h}+\frac{6 \gamma  f^2 h}{u}+\frac{3}{2} \gamma  f u f' h'-\frac{9}{2} \gamma  f h f'+\frac{h f'}{2}+\frac{f h'}{2}-\frac{2 f h}{u})\notag\\-i \gamma  f k^2 u \omega  h_{tx}'=0\label{8}.
\end{gather}
For simplicity, we eliminate $h_{tx}$ (including its derivatives terms) and define a new quantity
\begin{gather}
\chi'\equiv\frac{u}{f}\psi\label{9}.
\end{gather}
Next, we illustrate how to simplify the above equations and eliminate $h_{tx}$. Combine (\ref{6}) with (\ref{8}) and solve the formula of $h_{tx}$ and $h_{tx}'$. Then substitute $h_{tx}'$  and (\ref{9}) into (\ref{7}). Thus the first equation is obtained. Then, take a radial derivative of (\ref{6}) and solve the formula for $h_{tx}''$. Continue to take a radial derivative of (\ref{8}). Then substituting $h_{tx}$, $h_{tx}'$, $h_{tx}''$ and (\ref{9}) into the equation that we obtained before, the second equation is obtained. The full equations are too complicated to be present here, we leave them in Appendix. The asymptotic behavior of the Maxwell field near the boundary\,($u\to0$) is
\begin{gather}
a_x=a_x^{(0)}+a_x^{(1)}u+\cdots.
\end{gather}
According to holographic dictionary, the source and expectation value for the dual current operator are determined by $a_x^{(0)}$ and $a_x^{(1)}$, respectively. Then, the conductivity can be read off from the above coefficients\,\cite{Hartnoll:0803}
\begin{gather}
\sigma(\omega)=-\frac{ia_x^{(1)}}{\omega a_x^{(0)}}.
\end{gather}
And, the DC conductivity can be achieved by taking the zero frequency limit
\begin{gather}
\sigma_{DC}=\lim_{\omega\to 0}\sigma(\omega).
\end{gather}

\begin{figure}[htbp]
  \vspace{0.3cm}
  \centering
  \includegraphics[width=0.55\textwidth]{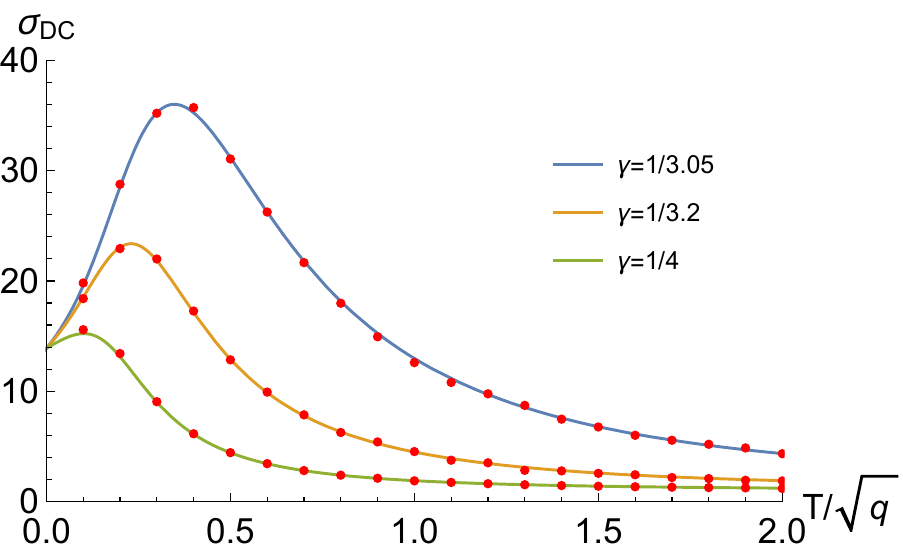}
  \caption{$\sigma_{DC}$ as the function of $T$ for different temperature. Blue, yellow and green lines are the analytical solutions obtained from (\ref{4})\,and red dots are the numerical solutions by solving (\ref{6}-\ref{8}).}\label{b}
  \vspace{0.3cm}
\end{figure}

In FIG.\ref{b}, we show a comparison between the numerical and analytical results for DC conductivity. From this figure, it can be easily seen that the numerical results are in good agreement with the analytical ones.

\begin{figure}[htbp]
  \vspace{0.3cm}
  \centering
  \includegraphics[width=0.45\textwidth]{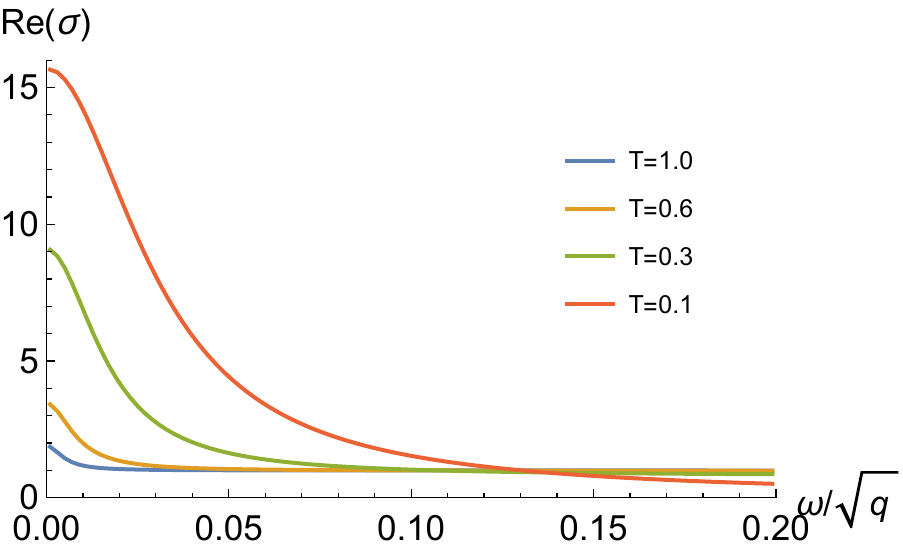}
  \includegraphics[width=0.45\textwidth]{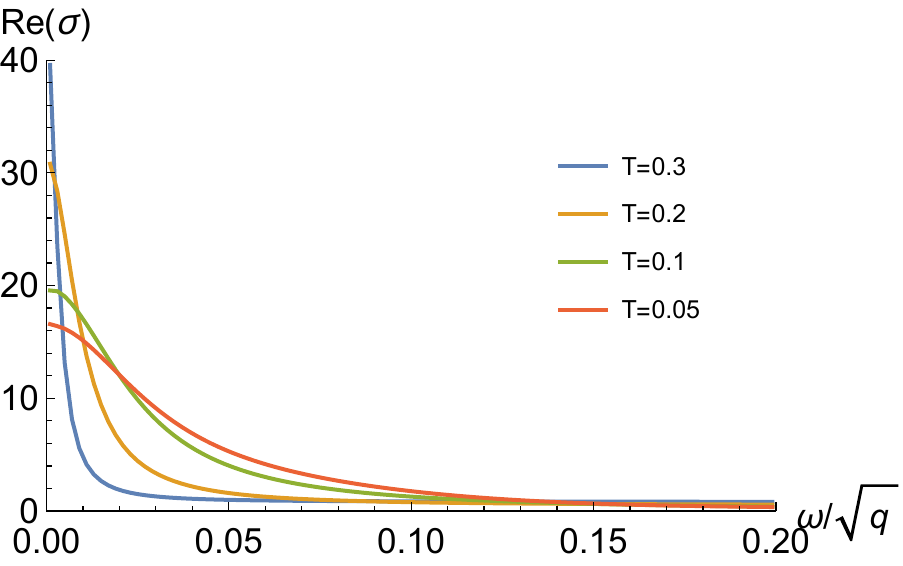}
  \includegraphics[width=0.45\textwidth]{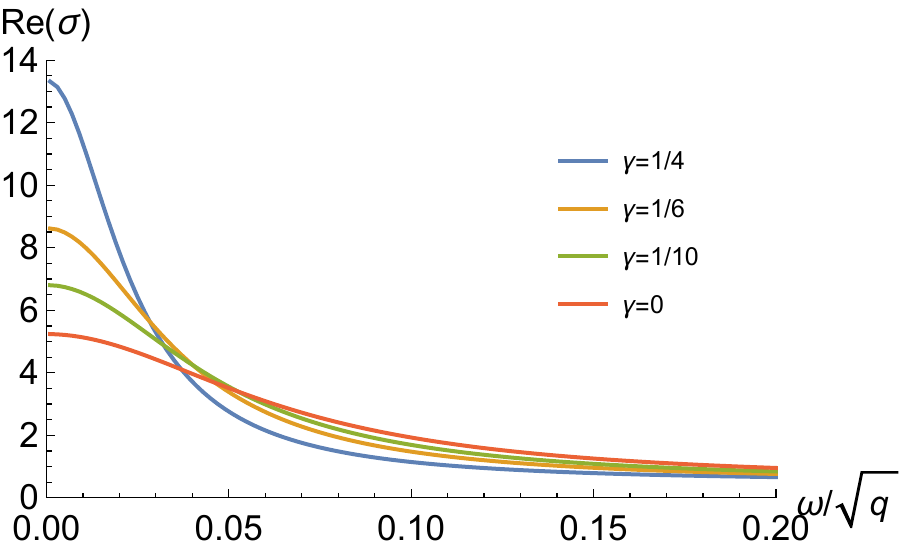}
  \caption{The dependance of the real part of AC conductivity on frequency for different temperatures or couplings. {\bf Top left}: we have fixed $\gamma$=1/4. {\bf Top right}: we have fixed $\gamma$=1/3.01. {\bf Bottom}: we have fixed $T$=0.2.}\label{c}
    \vspace{0.3cm}
\end{figure}

The numerical results for the real part of the AC conductivity have been displayed in FIG.\ref{c}. In the high frequency limit, the real parts of the AC conductivity are close to a non-zero constant which is an universal feature of the UV CFT fixed point . In the top left figure of FIG.\ref{c}, we find that the real part of the AC the conductivity always declines with the increase of the temperature at zero frequency, which is a typical behavior of metallic phase. While, in the top right figure of FIG.\ref{c}, the real part of the AC conductivity increases with the increase of the temperature at zero frequency, which seems insulator like. However, we find that the Drude peak never turns into an off-axis peak in the low frequency region like in massive gravities \cite{Baggioli:1411}.\footnote{Such peaks can be realized by considering some higher derivative terms which make the UV behavior of the axions different from the canonical case \cite{Alberte:1711,WJL:1808}. However, in our model, since $G^{\mu\nu}\propto g^{\mu\nu}$ in the UV, the Horndeski term does not change the UV expansion of the axions at all.} This in general, happens in many semiconducting systems whose DC conductivity obeys $d\sigma_{DC}/dT>0$ but still satisfies the Drude formula.  For more details, one refers to \cite{SmB6,Laurita:1709}.

\begin{figure}[htbp]
  \vspace{0.3cm}
  \centering
  \includegraphics[width=0.43\textwidth]{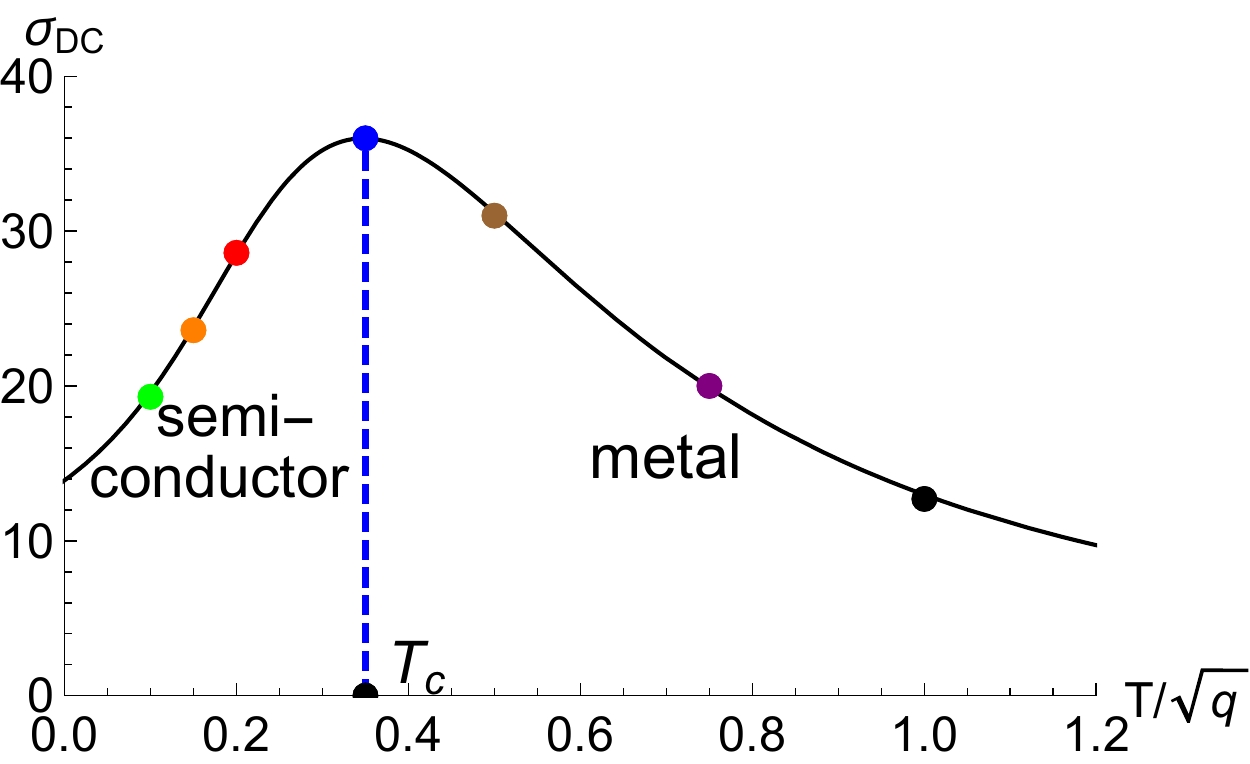}
  \includegraphics[width=0.50\textwidth]{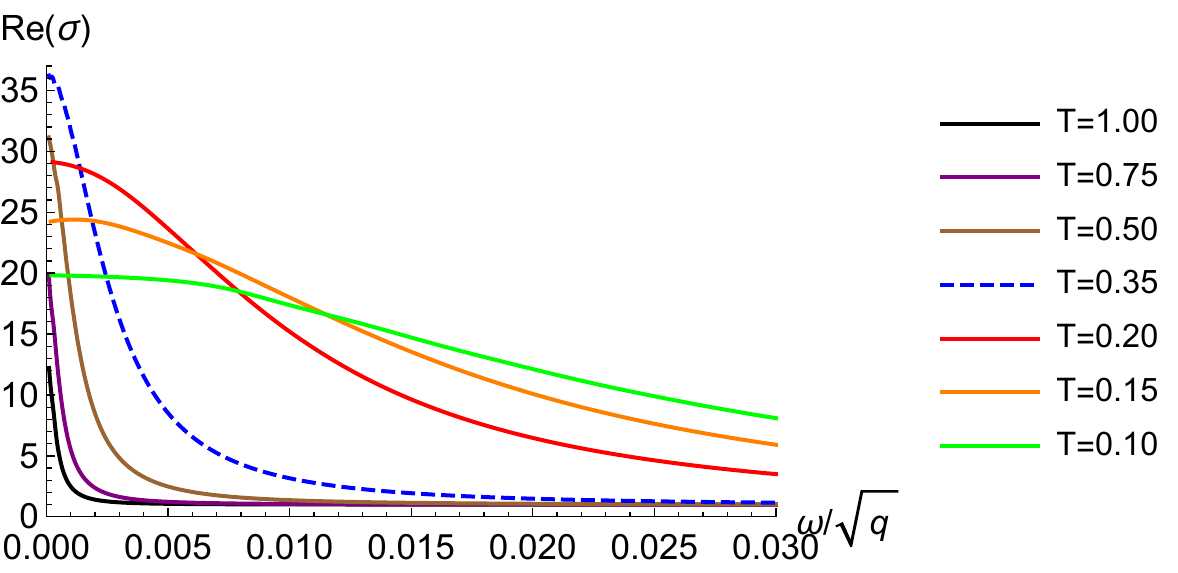}
  \caption{{\bf Left}: We have set $\gamma=1/3.05$ and the transition temperature $T_c$ is 0.35. {\bf Right}: The real part of the AC conductivity as the function of $\omega$.}\label{d}
    \vspace{0.3cm}
\end{figure}

In order to study the AC conductivity at both sides of the transition point, we choose a special coupling $\gamma=1/3.05$ and the transition temperature $T_c$ is 0.35 at this moment. The DC conductivity as the function of $T$ is shown in FIG.\ref{d}. The real parts of the AC conductivity are also shown in FIG.\ref{d}. The behaviors of AC conductivity imply this holographic model could provide a mechanism of semiconductor-metal transition driven by the Horndeski coupling term.

\begin{figure}[htbp]
  \centering
  \includegraphics[width=0.80\textwidth]{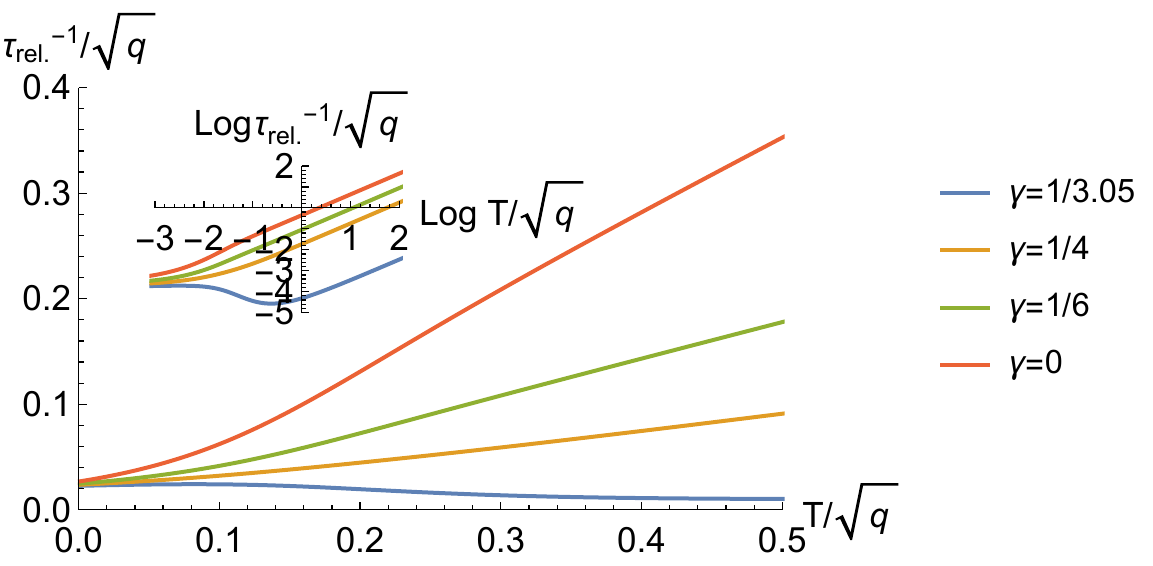}
  \caption{Relaxation rate $\tau_{rel.}^{-1}$ and its logarithmic form as the function of $T$ for different couplings. We have set $q=1$ here.}\label{e}
\end{figure}

\begin{figure}
  \centering
  \includegraphics[width=4.5cm]{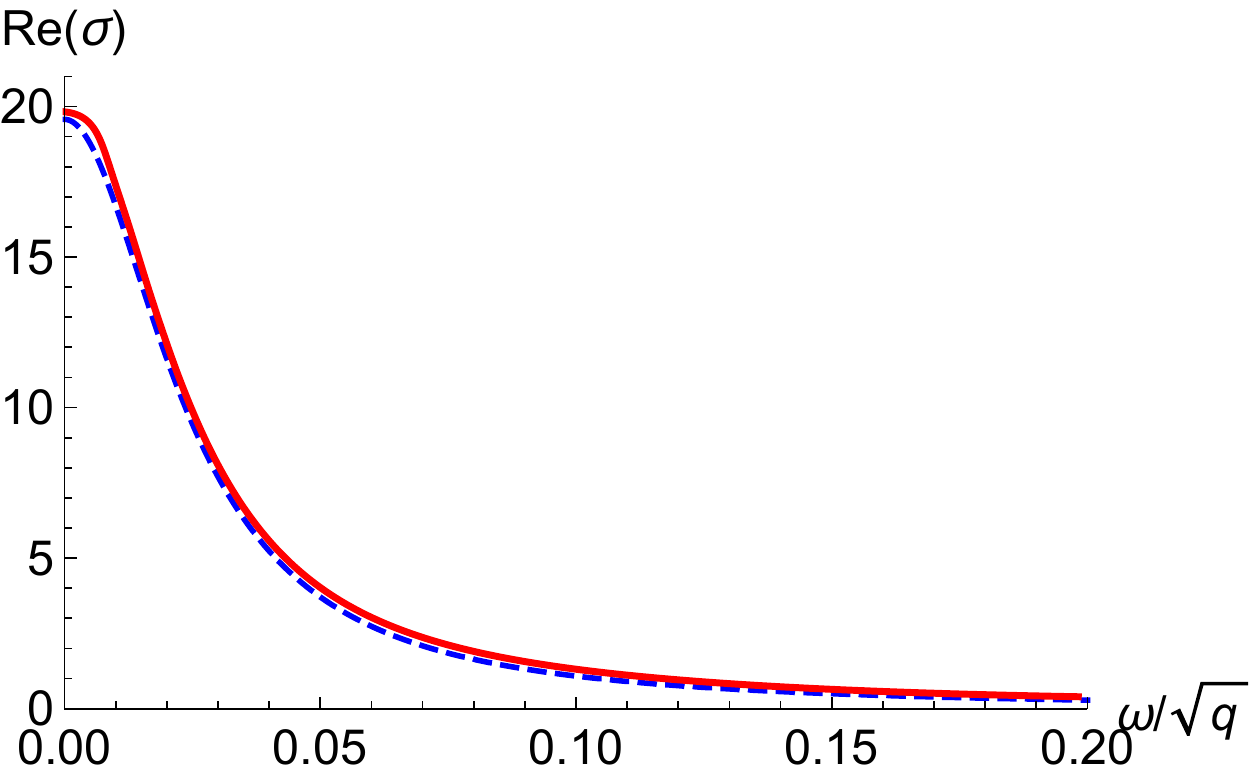}\hspace{3mm}
  \includegraphics[width=4.5cm]{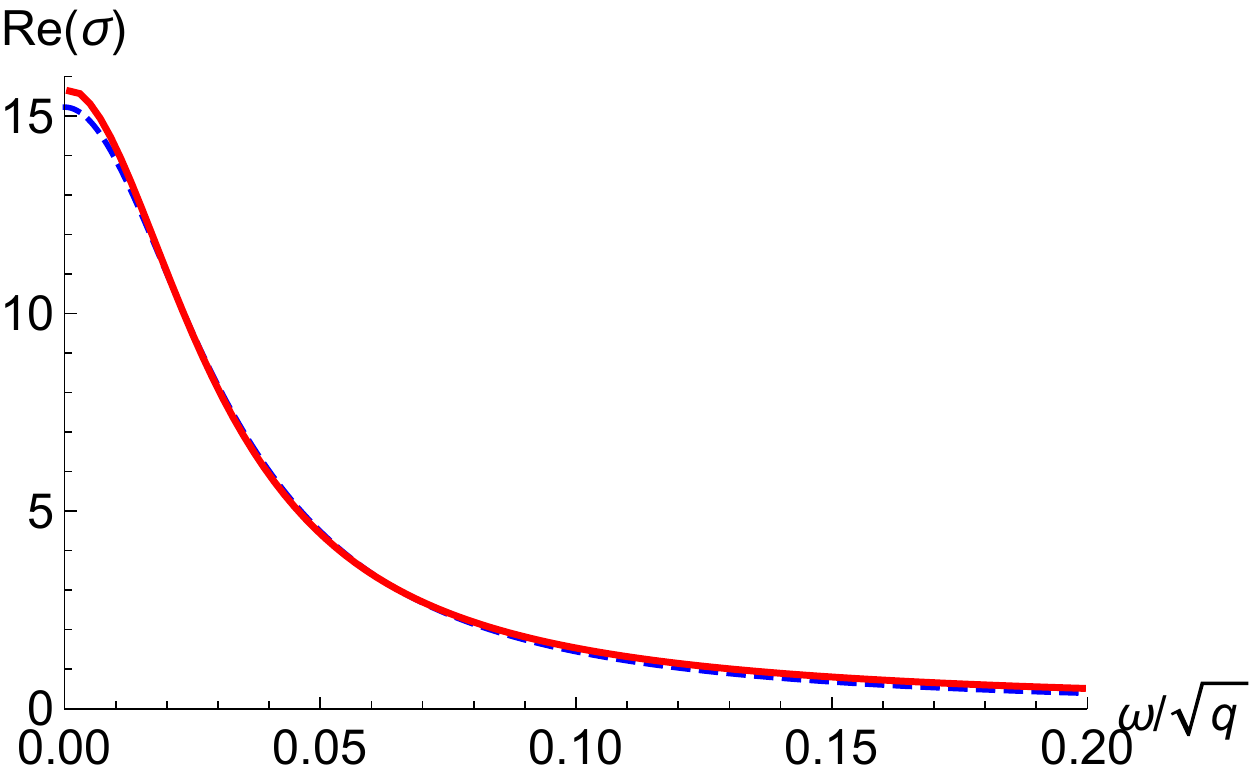}\hspace{3mm}
  \includegraphics[width=4.5cm]{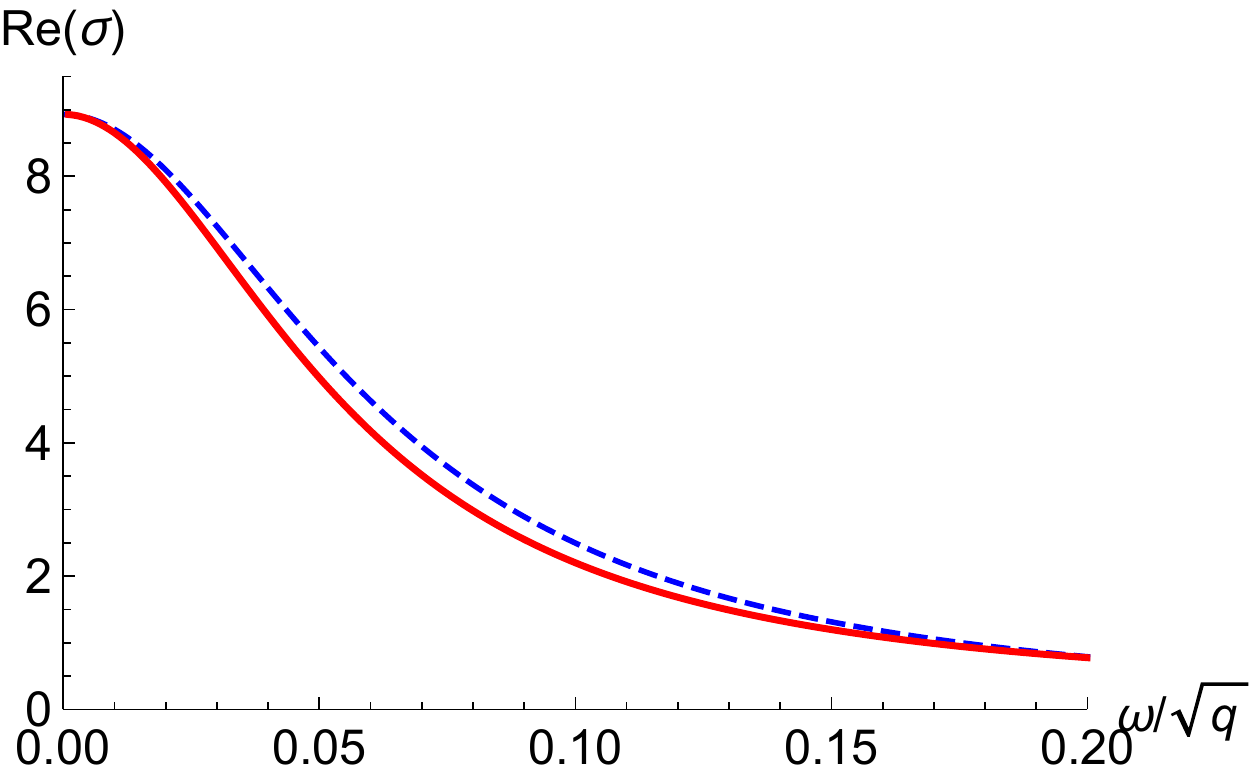}
  \includegraphics[width=4.5cm]{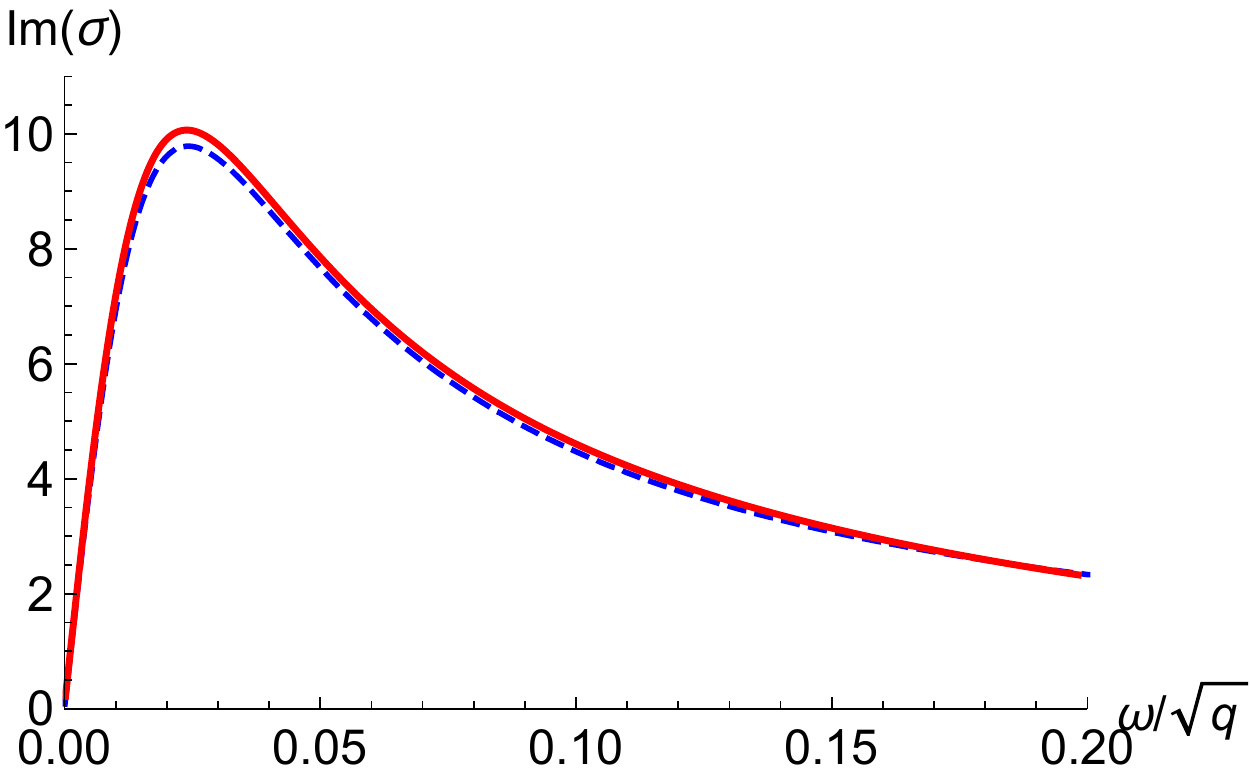}\hspace{3mm}
  \includegraphics[width=4.5cm]{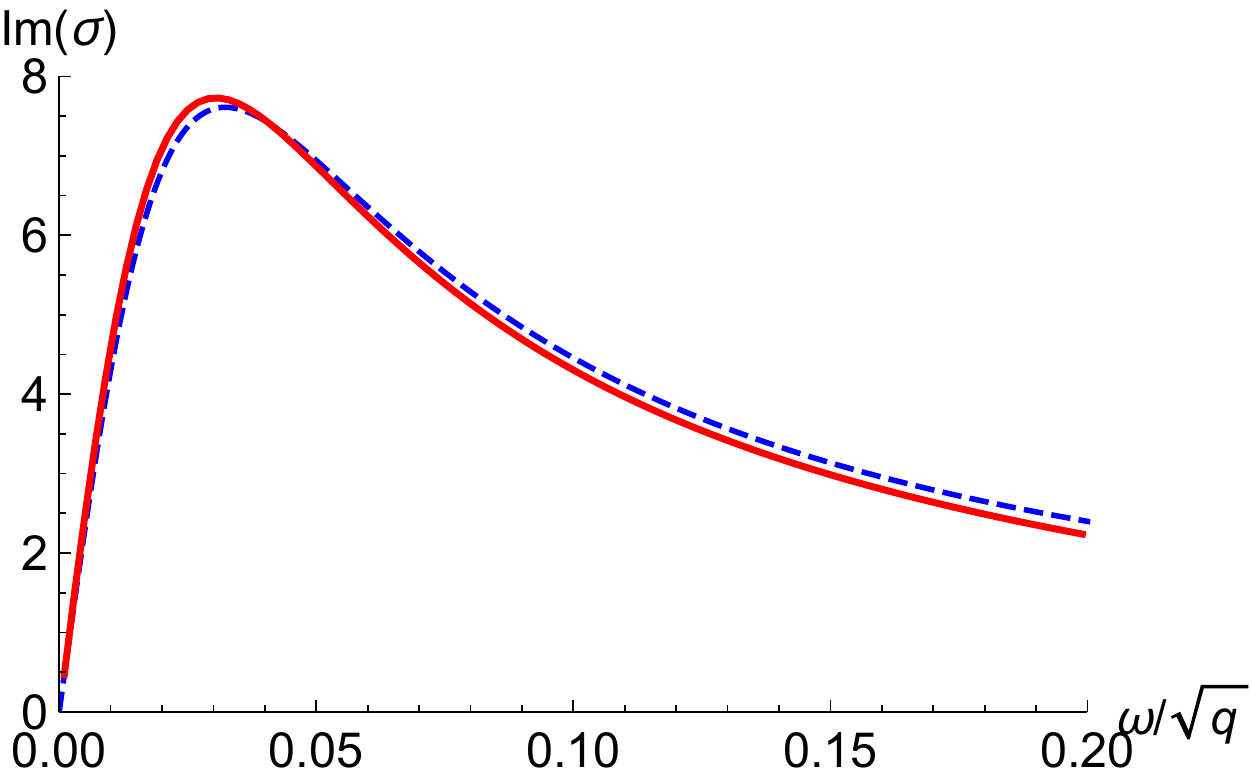}\hspace{3mm}
  \includegraphics[width=4.5cm]{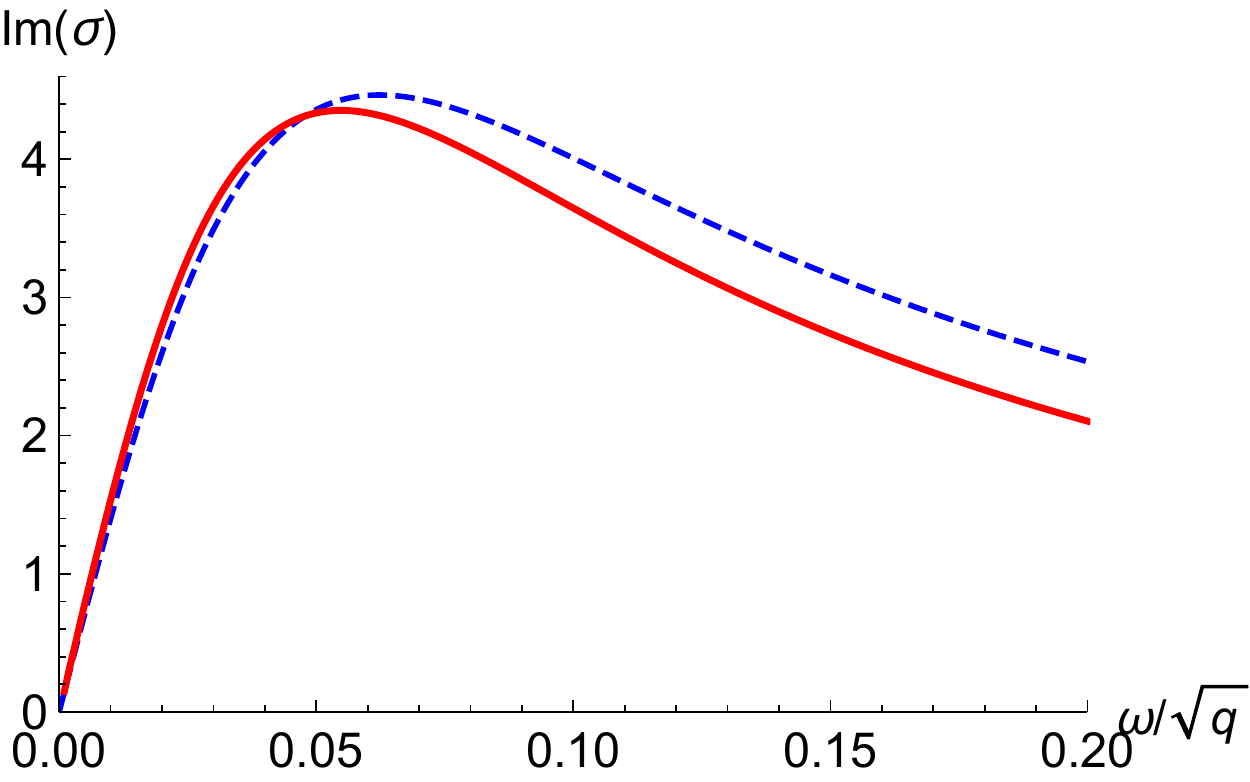}
  \caption{The real (top) and imaginary (bottom) parts of the AC conductivity for different $\gamma$'s. {\bf Left}: $\gamma=1/3.05$; {\bf Center}: $\gamma=1/4$; {\bf Right}: $\gamma=0$. Red solid lines are numerical solutions and blue dashed lines are results fitted by using Drude formula. Here, we have set $T=0.1$.}\label{f}
\end{figure}

Next, we fit above numerical results by using the Drude formula which is given by
\begin{equation}
\sigma(\omega)=\frac{\sigma_{DC}}{1-i\omega\tau_{rel.}}\label{10},
\end{equation}
where $\tau_{rel.}$ is the relaxation time. In general, the Drude formula only works for slow momentum relaxation. We then assume that the parameter $k$ is small. In this case $\tau_{rel.}$ can be calculated in a hydrodynamical approach. The inverse of the relaxation time is also called the relaxation rate. And, it takes the following form \cite{Davison:1306}
\begin{equation}
\tau_{rel.}^{-1}=\frac{sM_h^2}{2\pi(sT+\mu q)}\label{11}.
\end{equation}
Note that, to derive this formula, the first law of thermodynamics $\epsilon+P=sT+\mu q$ should be applied. The entropy density $s$ is given by\,\cite{WJL:1705}
\begin{equation}
s=\frac{4\pi}{u_h^2}(1-\gamma\frac{k^2 u_h^2}{2})\label{12}.
\end{equation}
Then, $\tau_{rel.}$ can be calculated by using (\ref{1}),\,(\ref{2}),\,(\ref{5}) and\,(\ref{12}). In the high $T$ limit, $\tau_{rel.}^{-1}$ becomes
\begin{equation}
\tau_{rel.}^{-1}\sim\frac{8\pi k^2(1-3 \gamma)}{9} T,
\end{equation}
 which implies that relaxation  rate is proportional to temperature. This is because the temperature is the only dimensional scale in this limit and the relaxation rate has the mass dimension. The figures of relaxation rate depending on the temperature are shown in FIG.\ref{e}.
The plots of the AC conductivity fitted by Drude formula are also shown in FIG.\ref{f}. From this figure, it can be seen the numerical AC conductivity results can be fitted well by Drude formula.

\section{Conclusion}
In this paper, we study the charge transport in a simple holographic Horndeski model with momentum relaxation. Firstly, we go over  the computation of the DC conductivity and show its relation with the  temperature. We find that when the Horndeski coupling is zero the DC conductivity declines monotonically with the increase of the temperature, behaving like a metal. While, when $\gamma$ reaches a critical value $1/3$, the conductivity always increases monotonically as the increase of the temperature, which approximately obeys the Steinhart-Hart equation for a semiconductor. When $\gamma$ is below the critical value, there always exists a transition point from the normal metal to the semiconductor like phase. Then, this holographic model may provide a metal-semiconductor transition that is driven by the Horndeski coupling.

To deepen our understanding on this transition, we further compute the AC conductivity by numerical method and obtain the dependance of the real part of AC conductivity on the frequency. We choose a special coupling to investigate the AC conductivity at both sides of the transition point. However, we do not find an off-axis peak in the low frequency region which usually happens in metal-insulator transitions.  Moreover, we also find that our numeric data of the AC conductivity fits well with the Drude formula for a semiconductor in slow relaxation cases.

\appendix
\section{Perturbation equations after simplification}
We have ever introduced that how to simplify the linearized perturbation equations (\ref{6})-(\ref{8}). Here, two equations after simplification will be presented. The first equation is
\begin{gather}
2 f h^2 \omega  a_x''(-\gamma  u f'+\gamma  f (\gamma  k^2 u^2+3)-1)+h^2 \omega  a_x' (2 f'+\gamma  f k^2 u) (-\gamma  u f'+\gamma  f (\gamma  k^2 u^2+3)-1)\notag\\+2 h \omega  a_x (f q^2 u^2 e^{\frac{1}{2} \gamma  k^2 u^2} (\gamma  u f'-3 \gamma  f+1)+\omega ^2 (-\gamma  u f'+\gamma  f (\gamma  k^2 u^2+3)-1))\notag\\+i q u \chi  e^{\frac{1}{4} \gamma  k^2 u^2} (\gamma  h u f' (3 h (\gamma  f-1)-\gamma  f u h')-2 h^2 (2 \gamma  f-1) (3 \gamma  f-1)+\gamma  f u (2 \gamma  f h u h''\notag\\-h' (\gamma  f u h'-3 \gamma  f h+h)))+2 i \gamma  f h q u^2 \chi ' e^{\frac{1}{4} \gamma  k^2 u^2} (\gamma  f u h'-3 \gamma  f h+h)=0.
\end{gather}
The second equation is
\begin{gather}
-i f k^2 q u^2 \omega  a_x e^{\frac{1}{4} \gamma  k^2 u^2} (\gamma  u (2 \gamma  f u f''+f' (-4 \gamma  u f'+\gamma  f (\gamma  k^2 u^2+8)-6))+\gamma  f (\gamma  k^2 u^2 (\gamma  f (\gamma  k^2 u^2-1)+1)\notag\\+6)-2)+2 i \gamma  f^2 k^2 q u^3 \omega  a_x' e^{\frac{1}{4} \gamma  k^2 u^2} (\gamma  u f'+\gamma  (-f) (\gamma  k^2 u^2+3)+1)+2 f^2 u \chi '' e^{\frac{1}{2} \gamma  k^2 u^2} (\gamma  u f'+\gamma  f (\gamma  k^2 u^2\notag\\-3)+1) (-\gamma  u f'+\gamma  f (\gamma  k^2 u^2+3)-1)+f u \chi ' e^{\frac{1}{2} \gamma  k^2 u^2} (\gamma ^3 f^3 k^2 u (\gamma  k^2 u^2 (3 \gamma  k^2 u^2+4)+9)-2 f' (\gamma  u f'+1)^2\notag\\+2 \gamma ^2 f^2 (3 u (f'' (\gamma  k^2 u^2+1)+k^2)+f' (2 \gamma  k^2 u^2 (\gamma  k^2 u^2+3)-15))-\gamma  f(\gamma  u f'+1) (2 u f''+f' (9 \gamma  k^2 u^2-16)\notag\\+3 k^2 u))+\chi  (f e^{\frac{1}{2} \gamma  k^2 u^2} (\gamma ^2 u^2 (f')^3(\gamma  k^2 u^2+6)+2 u (\gamma  f u (f^{(3)}(u) (\gamma  f (\gamma  k^2 u^2+3)-1)+\gamma  u f'' (f''\notag\\+k^2 (\gamma  f (2 \gamma  k^2 u^2+7)-3)))+k^2 (\gamma  f (\gamma  f (\gamma ^2 k^4 u^4 (4 \gamma  f-1)+2 \gamma  k^2 u^2 (3 \gamma  f-1)+9)-6)+1))\notag\\-2 \gamma  u (f')^2 (\gamma  u^2 f''+\gamma  f (\gamma  k^2 u^2 (\gamma  k^2 u^2+9)+6)-3 \gamma  k^2 u^2-4)+f' (\gamma (\gamma  f^2 (\gamma  k^2 u^2 (\gamma  k^2 u^2 (\gamma  k^2 u^2+2)\notag\\+37)+6)-2 f (\gamma  k^2 u^2 (\gamma  k^2 u^2+16)+4)+7 k^2 u^2)-2 \gamma  u^2 (\gamma  f u f^{(3)}(u)+f'' (\gamma  f (2 \gamma  k^2 u^2+1)+1))+2))\notag\\-2 u \omega ^2 (\gamma  u f'+\gamma  (-f) (\gamma  k^2 u^2+3)+1)^2)=0.
\end{gather}
\acknowledgments

We would like to thank M. Baggioli and Y. Liu for helpful discussions!  X. J. Wang and W. J. Li are supported by NFSC Grant No. 11905024 and DUT19LK20. H. S. Liu is supported in part by NSFC  Grants No. 11475148 and No. 11675144. X. J. Wang also would like to thank Y. Ling and M. H. Wu for the stimulating discussions and warm hospitality during the completion of part of this work.

\appendix

\end{document}